\documentclass[a4paper,aps,pre,onecolumn,groupedaddress,showkeys,floats,
floatats,floatfix]{revtex4}
\usepackage[utf8]{inputenc}
\usepackage[english]{babel}
\usepackage{graphicx}
\usepackage{dcolumn} 
\usepackage{bm}
\usepackage{natbib}
\usepackage{latexsym}
\usepackage{mathrsfs}
\usepackage{amssymb}
\usepackage{amsmath}
\usepackage{amscd}
\usepackage{color}
\usepackage{pifont}
\usepackage{pstricks,pst-node,pst-text,pst-3d}
\usepackage{verbatim}
\usepackage{ulem}
\usepackage[T1]{fontenc}
\usepackage{hyperref}
\hypersetup{
  colorlinks   = true, %colors links instead of ugly boxes
  urlcolor     = black, %color for external hyperlinks
  linkcolor    = red, %color of internal links
  citecolor    = blue %color of citations
}
\newrgbcolor{Red}{1.0 0.0 1.0}
\begin{document}
\title{Optimizing thermally affected ratchet currents using periodic
perturbations}
\author{Rafael M.~da Silva$^{1}$}
\email{rmarques@fisica.ufpr.br}

\author{Cesar Manchein$^{2}$}
\email{cesar.manchein@udesc.br}
\author{Marcus W.~Beims$^{1}$}
\email{mbeims@fisica.ufpr.br}
\affiliation{$^1$Departamento de F\'\i sica, Universidade Federal do Paran\'a, 
81531-980 Curitiba, PR, Brazil}
\affiliation{$^2$Departamento de F\'\i sica, Universidade do Estado de Santa 
Catarina, 89219-710 Joinville, SC, Brazil} 
\date{\today}
\begin{abstract}
  The purpose of the present work is to apply a recently proposed
  methodology to enlarge parameter domains for which optimal ratchet
  currents (RCs) are obtained. This task is performed by adding a
  suitable periodic perturbation $F_j$ on a Ratchet mapping and the 
  procedure consists in multiplying a particular class of {\it generic}
  Isoperiodic Stable Structures (ISSs), since the existence of non-zero RCs is
  directly related to the occurrence of stable domains. By proliferating
  the ISSs, it is possible to: (i) postpone thermal effects that usually  
  increase the chaotic domain and (ii) demonstrate, by using a
  quantitative analysis, that the area which provides optimal RCs in the
  two-dimensional parameter space can be enlarged about $78\%$. In
  addition, for some specific parameter combinations, non-zero RCs can
  be induced through the birth of a new attractor, which moves away as
  the strength of $F_j$ increases. Clearly, the methodology applied to
  the ratchet mapping is an efficient way to deal with unavoidable  
  thermal effects and its consequent undesirable dynamics, specially
  in experimental setups. As a second general remark, we conclude that
  our main findings can be extended to issues related to transport
  problems.    
\end{abstract}
%
%\pacs{05.45.Ac,05.45.Pq}
\maketitle
%===============================================================================

%%%%%%%%%%%%%%%%%%%%%%%%%%%%%%%%%%%%%%%%%%%%%%%%%%%%%%%%%%%%%%%
\section{Introduction}
\label{intro}
%%%%%%%%%%%%%%%%%%%%%%%%%%%%%%%%%%%%%%%%%%%%%%%%%%%%%%%%%%%%%%%

The transport phenomenon occurs in a diversity of realistic
situations, ranging  from physical \cite{B718850A}, chemical 
\cite{JCP2004}, to biological systems \cite{PhysRevLett.104.168102}.
In this context, an experimental and theoretical property of pronounced
relevance is the ``ratchet effect'', which consists in the transport
of particles with a preferred direction in systems without a net
driving force, or even against a small applied bias
\cite{AstumianScience,RevModPhys.69.1269,Reimann200257}. To achieve
the ratchet effect, spatio-temporal symmetries must be broken in the
system \cite{PhysRevE.75.066213,CHEN2014225}. A variety of models have been
proposed and experimentally performed to exemplify the unbiased
directed motion as, for example, solid state and superconducting
devices  \cite{RevModPhys.81.387} and cold atoms in optical lattices  
\cite{RevModPhys.81.387,PhysRevLett.94.164101, KOHLER2005379}, to
mention a few. 
 
To describe the transport of particles in nature and for actual
technological applications, where smaller and smaller scales becomes
relevant, intrinsic noise and thermal effects can develop into the
dominant processes. To properly understand such thermal effects on the
ratchet currents (RCs), we come back to the concept of {\it generic 
Isoperiodic Stable Structures} (ISSs), which are Lyapunov stable domains 
found in two-dimensional parameter space (named here as parameter space, 
for simplicity) of many dynamical systems \cite{Gallas93,
PhysRevLett.95.143905,Bonatto505}. It was shown recently that, in the 
absence of temperature, optimal RCs are deeply related with the
existence of the ISSs \cite{cmab-1062341012011,Celestino2014139}
and, that for increasing thermal effects, ISSs begin to be destroyed  
from their borders and periodic regions are replaced by chaotic
domains with smaller RCs \cite{mcb-1101141022013,ham17}.

As discussed above, since thermal effects cannot be ignored, it is
very desirable to propose and perform procedures which increase the
parameter ranges of operation and efficiency of experimental
devices. With this purpose in mind, an efficient way to avoid the
degradation of optimal RCs is to enlarge the area occupied by the
ISSs, parametric domains where non-zero RCs can be found. In a 
similar way from that used in the generation of multistability 
\cite{PhysRevE.64.036223} or duplication of ISSs \cite{MEDEIROS2010} in  
continuous dynamical systems through weak periodic perturbations, a
recently developed protocol shows that it is possible to generate
multiple attractors and ISSs in phase and parameter spaces,
respectively, by adding a periodic external parameter $F_j (j=1,2,\ldots,k)$ 
on discrete dynamical systems  \cite{rafael17-1,rafael17-2}. Then, increasing 
the strength of $F_j$, multiple  identical copies of the ISSs start to separate 
from each other, enlarging the available stable domains in phase and parameter 
spaces. In this scenario, such methodology can be successfully applied in the 
Ratchet Mapping (RM) to postpone thermal effects, keeping a
considerable area of the parameter space leading to optimal
RCs. Besides that, another remarkable related phenomenon consists on
the activation of RC by temperature in classical systems
\cite{mcb-1101141022013} and also by quantum fluctuations in  
quantum systems \cite{PhysRevE.91.052908}. Despite the thermal
activation of RCs  in specific regions of the parameter space, optimal
RCs inside the ISSs are still destroyed and, in this work, we
demonstrate that it is possible to {\it induce} the creation of
regions with non-zero RCs keeping the vicinity intact. Investigations
about applications of ratchet effects are still very active
\cite{HUANG2018771,ding18,bram17,brox17,erbas17,gall17,WANG201784,hans16, 
budkin16,koenig16}, and this is what motivates us to apply our
methodology to this system.  

This work is organized as follows: In Sec.~\ref{meth}, the method is
explained and the model is presented as well the influence of the
periodic external perturbation $F_j$ on its dynamics. In
Sec.~\ref{curr}, we study how the multiplication of ISSs can enlarge
the region with non-zero RCs  and how to obtain the optimal value for
$F_j$ that maximizes the required results. Sec.~\ref{induce} shows the
induction of RCs by using temperature and external forces, and
finally, in Sec.~\ref{conc}, we summarize our main results.

%%%%%%%%%%%%%%%%%%%%%%%%%%%%%%%%%%%%%%%%%%%%%%%%%%%%%%%%%%%%%%%
\section{Theoretical explanation about the proliferation of stable
  domains} 
\label{meth}
%%%%%%%%%%%%%%%%%%%%%%%%%%%%%%%%%%%%%%%%%%%%%%%%%%%%%%%%%%%%%%%

Recently, using one-dimensional quadratic maps modified by the
addition of a $k$-periodic external perturbation $F$, it was shown
that it is possible to create multiple attractors and move apart
independent bifurcation diagrams by increasing the value of $F$
\cite{rafael17-1}. With shifted bifurcation diagrams, the parameter
range for which occurs periodic dynamics is enlarged due to the
appearance of extra-stable motion through saddle-node bifurcations. By
applying this procedure in two-dimensional maps, identical
multioverlapping ISSs are created and, increasing the strength of
$F$, it is possible to separate them, given rise to the {\it
  proliferation of stability} in the  parameter space
\cite{rafael17-2}. These procedures are general and should be
applied to any high dimensional nonlinear discrete-time system.  

When a $k$-periodic external parameter $F_j$, with $j=1,2,\ldots,k$,
changes the dynamics of a two-dimensional map $\mathbf{T}_n$ by a
different value on each time iteration, it is necessary to analyze the
map $\mathbf{T}_{n+k}^{(c)}$ composed by $k$ iterations of
$\mathbf{T}_n$ to understand how the proliferation of stable domains 
occurs. For instance, consider the map
$\mathbf{T}_n=f(\boldsymbol{x}_n,\alpha,F_j)$, with $\boldsymbol{x}_n$
being the variables and $\alpha$ representing all parameters
involved. In this case, the composed map $\mathbf{T}^{(c)}_{n+k}$ is
obtained by applying $\mathbf{T}_n$ $k$-times, employing all possible
values for $F_j$ as follows: 
\begin{equation}
  \label{composed}
  \mathbf{T}^{(c)}_{n+k} = f(\boldsymbol{x}_{n+k-1},\alpha,F_k) \circ\ldots\circ 
  f(\boldsymbol{x}_{n+1},\alpha,F_2) \circ
  f(\boldsymbol{x}^{(c)}_n,\alpha,F_1).
\end{equation}

After $k$ iterations of $\mathbf{T}_n$ we have one iteration of the 
composed map (represented by the superscript $c$). States between 
$\boldsymbol{x}^{(c)}_n$ and $\boldsymbol{x}^{(c)}_{n+k}$, namely
$\boldsymbol{x}_{n+k-1}, \ldots, \boldsymbol{x}_{n+1}$, are called
intermediate states. For the simplest case $k=2$, we can use $F_1=+F$,
$F_2=-F$ and, from Eq.~(\ref{composed}), we obtain
$\mathbf{T}^{(c)}_{n+2}=f(\boldsymbol{x}_{n+1}, \alpha,- F) \circ
f(\boldsymbol{x}^{(c)}_n,\alpha,+F)$. Therefore, after $k=2$
iteration of the map  $\mathbf{T}_n$ we obtain one iteration of 
$\mathbf{T}^{(c)}_{n+2}$.  We define $m$ as the period inside ISSs for 
$F=0$ and $m_F$ the period for $F\ne 0$. To multiply ISSs in the
parameter space, some rules that relate $k$ and the period $m$ of a
specific ISS, must be obeyed. When the ratio $\omega=m/k$ is an
integer, $k$ attractors with same period in phase space and
$k$-identical ISSs in the parameter space are generated. The multiplied
ISSs have period $m_F=m$, equal shape and equal Lyapunov stability
conditions. On the other hand, when the ratio $\omega$ is not an
integer, the number of attractors in phase space and ISSs in parameter
space remain unaltered. In this case, the new orbital period is $m_F =
km$ and the ISS is broken apart \cite{rafael17-2}. 

%%%%%%%%%%%%%%%%%%%%%%%%%%%%%%%%%%%%%%%%%%%%%%%%%%%%%%%%%%%%%%%
\subsection{The discrete-time mathematical model}
\label{model}
%%%%%%%%%%%%%%%%%%%%%%%%%%%%%%%%%%%%%%%%%%%%%%%%%%%%%%%%%%%%%%%
In order to show how to multiply stable domains and, consequently,
regions in the parameter space with non-zero RCs, we use the {\it
  Ratchet Mapping} (RM) which presents all essential features
regarding unbiased current
\cite{PhysRevLett.94.164101,PhysRevLett.99.244101,PhysRevE.79.026212}: 
\renewcommand{\arraystretch}{1.2}
\begin{equation}
\label{map-ratchet}
\begin{array}{ll}
p_{n+1} = \gamma p_{n} + K [\mathrm{sin}(x_n) + a \hspace{0.05cm} 
\mathrm{sin}(2 x_n + \phi)] + \sqrt{2(1-\gamma)k_{\mbox {\tiny B}} T} 
\hspace{0.1cm} \zeta + F_j, \\
x_{n+1} = x_{n} + p_{n+1}, \\
\end{array}
\end{equation}
\renewcommand{\arraystretch}{1}
where $F_j$ is a $k$-periodic perturbation and $K$ the nonlinearity
parameter. This model describes the dynamics of a particle that moves
in an asymmetric potential in the $x$ direction with $x\in
(-\infty,+\infty)$, $p$ is the conjugate momentum of $x$ and $\gamma
\in [0,1]$ is the dissipation. The limit $\gamma=0$ is the overdamped
case and the conservative limit corresponds to $\gamma=1$, since
$|\text{Det} \hspace{0.05cm} \mathbf{J}_n| = \gamma$, where
$\mathbf{J}_n$ is the Jacobian matrix for the RM
(\ref{map-ratchet}). The spacial symmetry is broken when $a \neq 0$
and $\phi \neq l \pi$, where $l$ is an integer. The term $\zeta$
represents the Gaussian noise with $\langle \zeta_n \rangle=0$ and
$\langle \zeta_{n_1} \zeta_{n_2} \rangle = \delta_{n_1n_2}$, while
$k_{\mbox {\tiny B}}=1$ is the Boltzmann constant and $T$ the
temperature. An importance  requirement for $F_j$ is that $\langle F_j
\rangle=0$ after $k$ iterations of the RM (\ref{map-ratchet}), since
the added external perturbation should not, to keep the ratchet
effect, generate directed motion.  
%
%%%%%%%%%%%%%%%%%%%%%%%%%%%%%%%%%%%%%%%%%%%%%%%%%%%%%%%%%%%%%%%%%%%%%%%%%
\begin{figure}[!b]
  \centering
  \includegraphics*[width=0.96\columnwidth]{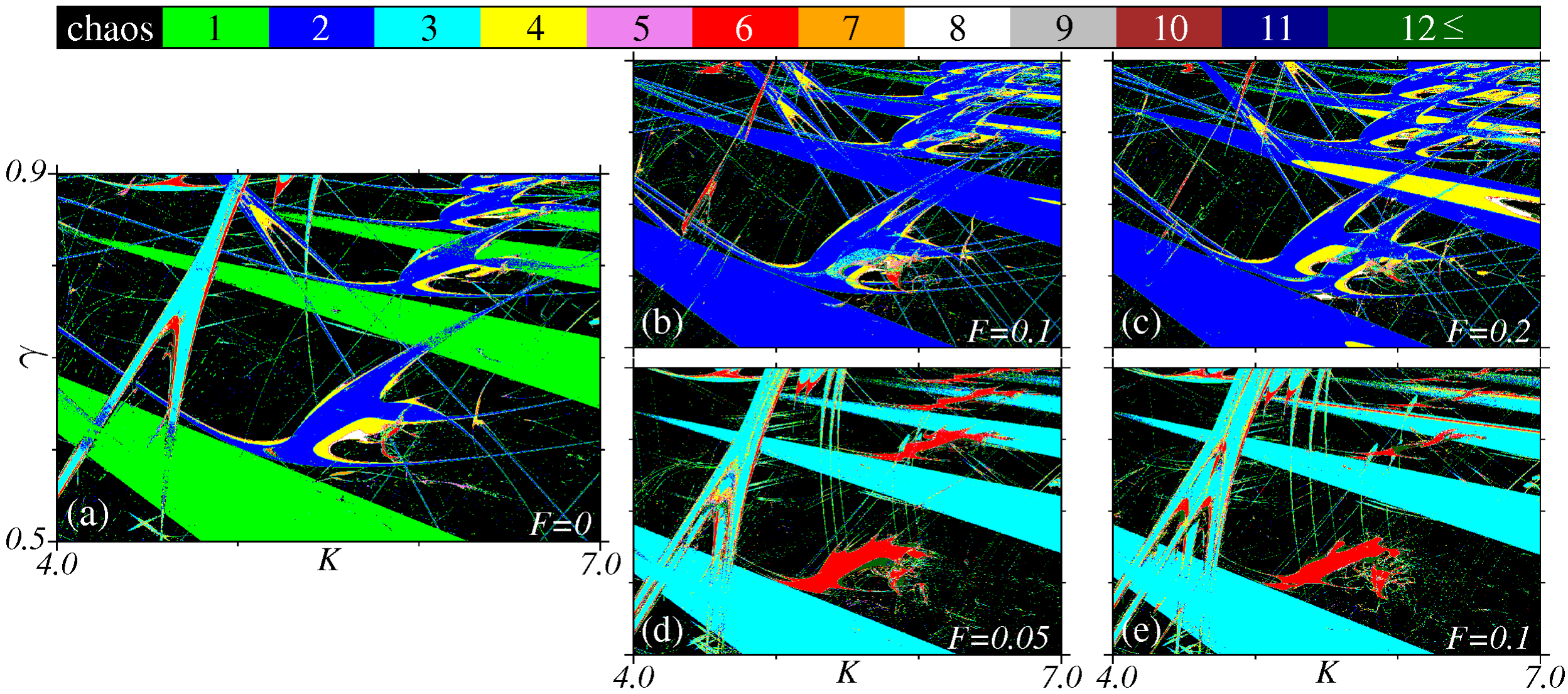}
  \caption{(Color online) Parameter space $(K,\gamma$) inside the interval 
  $(K_{\mbox{\tiny min}}, K_{\mbox{\tiny max}})=(4.0,7.0)$ and 
  $(\gamma_{\mbox{\tiny min}},\gamma_{\mbox{\tiny max}})=(0.5,0.9)$ for the map 
  (\ref{map-ratchet}) with $T=0$. The orbital period was counting after a 
  transient time of $10^7$ iterations. In (a), the case $F=0$ is displayed 
  (already published in \cite{cmab-1062341012011}), in (b) and (c), a per-2 
  external perturbation was used with $F=0.1$ and $F=0.2$,
  respectively, and the case with a per-3 external perturbation is
  shown in (d) ($F=0.05$) and (e) ($F=0.1$). The initial condition
  (IC) used is $(x_0,y_0)=(0.1,0.1)$.}  
  \label{rat-per}
\end{figure}
%%%%%%%%%%%%%%%%%%%%%%%%%%%%%%%%%%%%%%%%%%%%%%%%%%%%%%%%%%%%%%%%%%%%%%%%%

Initially for the noiseless case ($T=0$), we study the parameter space
($K,\gamma$) with $a=0.5$ and $\phi=\pi/2$, using different sequences
and values for $F_j$ and counting the orbital period for only one
  initial condition $(x_0,p_0)=(0.1,0.1)$ restarted for every parameter
  combinations. For each parameter pair ($K,\gamma$) we discard a   
  transient time of $10^7$ iterations to avoid any influence of 
  transient effects. For reference, in
Fig.~\ref{rat-per}(a) we plot the case $F=0$ which was already
published before \cite{cmab-1062341012011}, but reproduced
here just for comparison with cases for which $F\neq 0$. In
Fig.~\ref{rat-per}(a) there is a chaotic region in black and ISSs with
periods $1$ (green), $2$ (blue) and $3$ (cyan). Using a per-$2$
external perturbation $+F,-F,+F,-F,\ldots$ with intensity $F=0.1$ we
obtain Fig.~\ref{rat-per}(b), where ISSs with even period are
duplicated, given rise to one more identical copy (see the blue
per-$2$ ISSs). On the other hand, stable domains with odd period are
not duplicated, however, their period become duplicated, namely $1 \to
2$ and $3 \to  6$. The ISS with per-3 is almost destroy by the
external perturbation. It is also important to emphasize that all
intermediate iterations are considered to count the
period. Fig.~\ref{rat-per}(c), obtained using $F=0.2$, clearly shows
that the intensity $F$ controls the separation between the ISSs. We
can also increase the period of $F$ from two to three using a sequence
$+F,0,-F,+F,0,-F,\ldots$. In this case, displayed in
Figs.~\ref{rat-per}(d) and \ref{rat-per}(e) for $F=0.05$ and $F=0.1$,
respectively, only ISSs with periods multiple of three are
multiplied. This can be seen by the triplicated cyan per-$3$ ISSs in
Figs.~\ref{rat-per}(d) and \ref{rat-per}(e). They follow the rule for
integers $\omega=m/k=3/3$, while other periods becomes $m_F=3m$. With
these results we can conclude that the RM (\ref{map-ratchet}) follows
the general rules for systems perturbed by a $k$-periodic external
parameter \cite{rafael17-1,rafael17-2}. We believe that this result
corroborates with the general character inherent to the methodology
applied here. 

%%%%%%%%%%%%%%%%%%%%%%%%%%%%%%%%%%%%%%%%%%%%%%%%%%%%%%%%%%%%%%%%%%%%%%%%%
\section{Optimizing regions with non-zero RC\lowercase{s}}
\label{curr}
%%%%%%%%%%%%%%%%%%%%%%%%%%%%%%%%%%%%%%%%%%%%%%%%%%%%%%%%%%%%%%%%%%%%%%%%%

%%%%%%%%%%%%%%%%%%%%%%%%%%%%%%%%%%%%%%%%%%%%%%%%%%%%%%%%%%%%%%%%%%%%%%%%%
\begin{figure}[!b]
  \centering
  \includegraphics*[width=1.0\columnwidth]{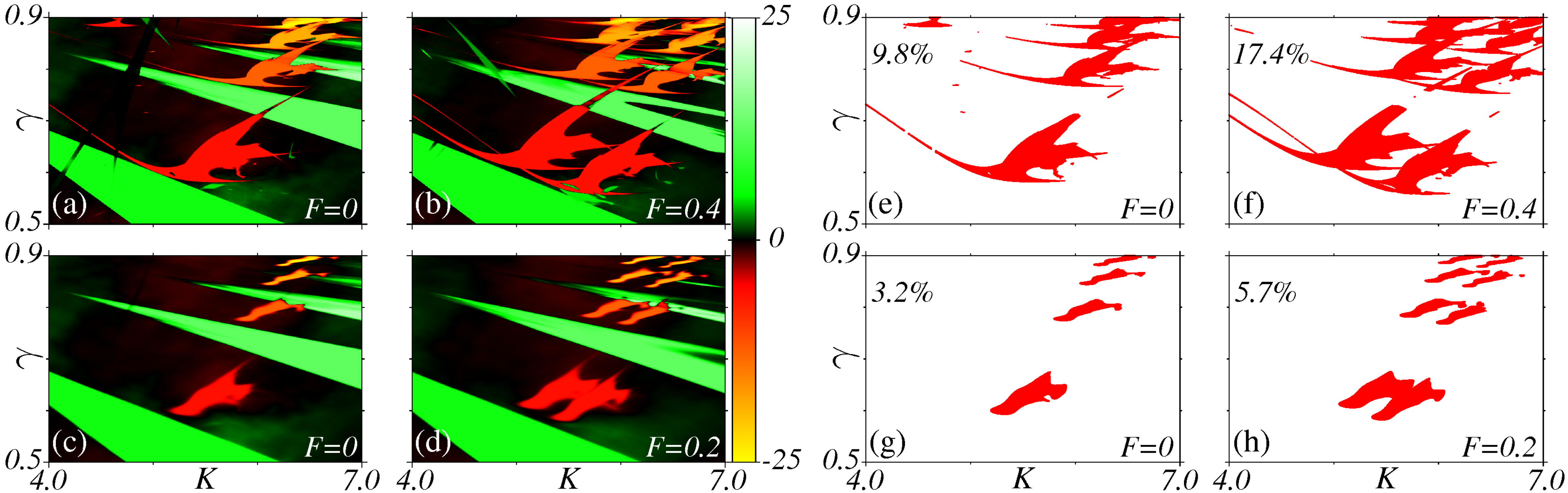}
  \caption{(Color online) RCs $\mathcal{J}$ plotted in the parameter
    space $(K,\gamma$) inside the interval
    $(K_{\mbox{\tiny min}},K_{\mbox{\tiny max}}) = (4.0,7.0)$ and
    $(\gamma_{\mbox{\tiny min}},\gamma_{\mbox{\tiny max}})=(0.5,0.9)$
    [see Fig.~3(a) in \cite{cmab-1062341012011}]. In (a)-(b) and  
  (c)-(d), the cases $T=10^{-5}$ and $T=10^{-3}$ are shown,
    respectively. In (e)-(h), only parameter combinations with
    $\mathcal{J}<-5$ are plotted for same temperature used in (a)-(d),
    showing the area occupied by these values of RCs. The intensity
    $F$ of the external force used in each case is described in the
    panel.} 
  \label{rat-cur}
\end{figure}
%%%%%%%%%%%%%%%%%%%%%%%%%%%%%%%%%%%%%%%%%%%%%%%%%%%%%%%%%%%%%%%%%%%%%%%%

The most relevant quantity for the RM, when it comes to transport
properties, is the {\it Ratchet Current} (RC), named as $\mathcal{J}$
and defined as a double average of momentum $p$:
\begin{equation}
  \label{current}
  \mathcal{J} = \dfrac{1}{M} \displaystyle \sum_{i=1}^M \left[\dfrac{1}{N} 
    \displaystyle \sum_{n=1}^N p_n^{(i)}  \right], 
\end{equation}
where $M$ is the number of initial conditions (ICs) and $N$ the total
time iteration. The ICs must be equally distributed between the
interval $(x_{\mbox{\tiny min}},x_{\mbox{\tiny max}})=(p_{\mbox{\tiny
    min}}, p_{\mbox{\tiny max}})=(-2\pi,2\pi)$, since the mean
of $x_0$ and $p_0$ should not establish a preferred direction. The
same interval used in the parameter spaces from Fig.~\ref{rat-per} is
displayed in Fig.~\ref{rat-cur}(a), but the RC $\mathcal{J}$ is
plotted and thermal effects are considered using $T=10^{-5}$. In this
case, black color represents RCs around zero and, when compared with
Fig.~\ref{rat-per}(a), it is possible to clearly see that regions with
zero RCs correspond to the chaotic dynamics. On the other hand, while
increasing positive RCs are found in regions with colors going from
green to white, negative increasing RCs are represented by red and
yellow colors. The case $T=10^{-5}$ is shown in Figs.~\ref{rat-cur}(a)
and \ref{rat-cur}(b) for $F=0$ and $F=0.4$, respectively, using the
same sequence $+F,-F,+F,-F, \ldots$. To obtain $\mathcal{J}$, $10^4$
ICs are considered and the time average was performed using
$5\times10^4$ iterations after a transient time with same number of
iterations. The transient time-interval discarded here is 
  long enough,  so the system~(\ref{map-ratchet}) already reached the
  asymptotic behavior. Figs.~\ref{rat-per}(a) and \ref{rat-cur}(a),
  both for $F=0$, show that ISSs are organized along a preferential
  direction [see ISSs with per-1   (green) and per-2 (blue) in
    Fig.~\ref{rat-per}(a)]. Along this direction, the absolute value
  of the RC $\mathcal{J}$ increases $2\pi$ inside ISSs of same period
  \cite{cmab-1062341012011}, as one can see in Fig.~\ref{rat-cur}(a), 
where per-$1$ ISSs tend to change colors from green to white and
per-$2$ ISSs from red to yellow. Besides that, we can observe that the
thermal noise tends to destroy the ISSs, always starting from their
antennas, decreasing the parameter combinations which generate
non-zero RCs. When Figs.~\ref{rat-cur}(a) and \ref{rat-cur}(b) are
compared, we clearly note that regions with negative current (red and
yellow) are duplicated, and therefore, enlarged when $F\neq 0$. This
occurs because a per-$2$ external perturbation $F_j$ was used in
Eq.~(\ref{map-ratchet}) and only ISSs with even periods are
duplicated. The per-$1$ ISSs with positive RCs tend to be destroyed
with increasing intensity of $F_j$ [see green ISSs in
  Fig.~\ref{rat-cur}(b)]. The per-$3$ ISS is almost destroyed for
$F=0.4$. Therefore the period and intensity of $F_j$ must be chosen
properly, depending on the desired aim. In Fig.~\ref{rat-cur}(b) the
intensity $F=0.4$ was used to induce a greater separation between the
ISSs, optimizing the enlargement of the area with negative RC.

\begin{figure}[!t]
  \centering
  \includegraphics*[width=0.42\columnwidth]{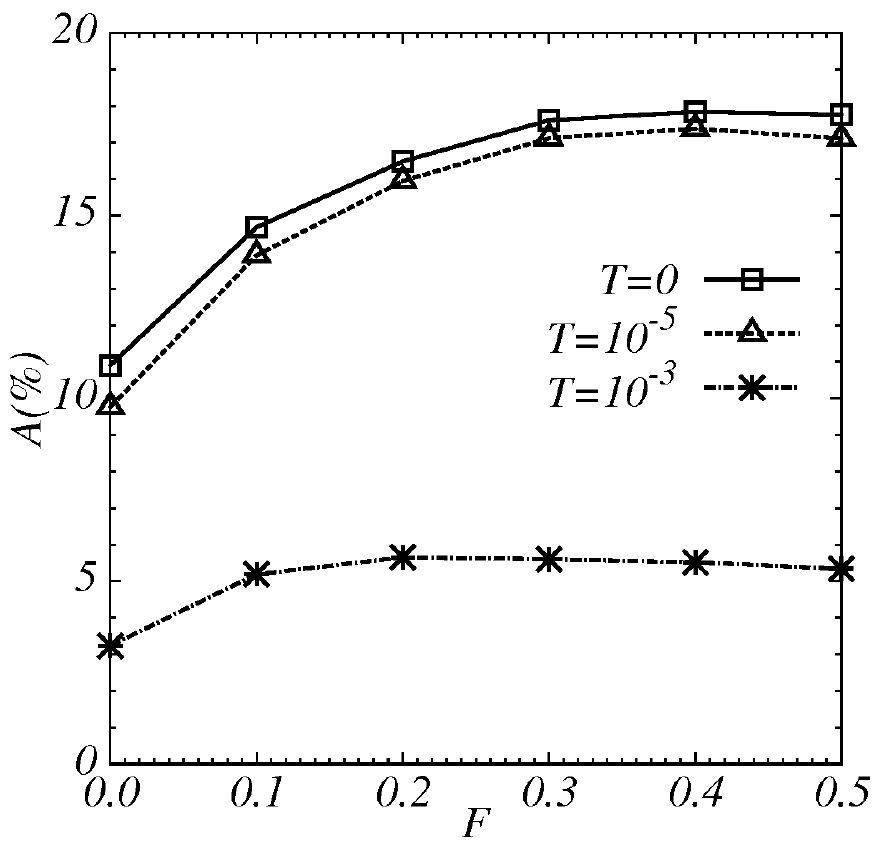}
  \caption{The percentage of occupied area $A(\%)$ of the parameter
    space $(K,\gamma)$ with $\mathcal{J}<-5$ as a function of $F$ for
    different values of temperature.}
  \label{percent}
\end{figure}

In Figs.~\ref{rat-cur}(c) and \ref{rat-cur}(d), the case for
$T=10^{-3}$ is displayed using $F=0$ and $F=0.2$, respectively,
showing that ISSs are resistant to reasonable values of
temperature. Increasing the thermal noise strength, a considerable
portion of the ISSs is destroyed, starting by their
borders. Consequently, the amount of regions in the parameter space
with approximately zero RCs tends to increase. Fig.~\ref{rat-cur}(d)
shows the case for $F=0.2$ with the same sequence $+F,-F,+F,-F,\ldots$
and the enlargement of the region with negative RCs is obtained,
showing that our methodology also works for larger
temperatures. Figs.~\ref{rat-cur}(e)-\ref{rat-cur}(h) display
$\mathcal{J}$ only for those parameter combinations which lead to
$\mathcal{J}<-5$, together with the corresponding percentage of area
occupied by these domains when compared to the entire parameter
space. These figures show that, for $F\neq 0$, the enlargement in the
area ($9.8\%\to 17.4\%$) for this specific value of RC represents a
gain of $77.5\%$ for $T=10^{-5}$. For $T=10^{-3}$, this gain is even
more significant, around $78.1\%$. The periodic perturbation
intensities, $F=0.4$ for the first value of $T$ and, $F=0.2$ for the
second one, were suitable chosen because they optimize the region with
RCs $\mathcal{J}<-5$. These values, redefined here as
$F^{\mbox{\tiny (opt)}}_T$, can be obtained from Fig.~\ref{percent},
which shows how the area in the parameter space with
$\mathcal{J}<-5$ changes as function of $F$.  
%
%%%%%%%%%%%%%%%%%%%%%%%%%%%%%%%%%%%%%%%%%%%%%%%%%%%%%%%%%%%%%%%%%%%%%%%%%
%%%%%%%%%%%%%%%%%%%%%%%%%%%%%%%%%%%%%%%%%%%%%%%%%%%%%%%%%%%%%%%%%%%%%%%%%

Another essential information obtained from the curves plotted in
Fig.~\ref{percent} is the percentual of occupied area in the parameter
space by parameter combinations that lead to non-zero RCs when the
thermal intensity is increased. For example, considering $F=0$ and
$T=0$, the occupied area with $\mathcal{J}<-5$ is equal
$10.9\%$. When comparing this value with $9.8\%$ for $T=10^{-5}$ and
$3.2\%$ for $T=10^{-3}$, this is a reduction of $10.1\%$ for the first
case and $70.6\%$ for the second one. On the other hand, if we compare
the occupied area for $F=0$ and $T=0$ ($10.9\%$) with the cases for
which optimal values of $F$ are used, {\it i.e.},
$F^{\mbox{\tiny (opt)}}_{10^{-5}}=0.4$ and $F^{\mbox{\tiny (opt)}}_{10^{-3}}=0.2$,
resulting in $17.4\%$ and $5.7\%$, respectively, we obtain a gain of  
$59.6\%$ even if $T=10^{-5}$, and a decrease of only $47.7\%$ for
$T=10^{-3}$ instead $70.6\%$ when $F=0$.

These results allow us to conclude that through of external
perturbations with a  suitable strength and period, it is possible to
postpone thermal effects keeping a considerable area of parameter
domains in the parameter space with a desired value of
$\mathcal{J}$. As optimal ratchet currents represent an imperative
issue usually taken into account in transport problems, we are
convinced about the importance of these results.   

%%%%%%%%%%%%%%%%%%%%%%%%%%%%%%%%%%%%%%%%%%%%%%%%%%%%%%%%%%%%%%%
\section{Activating RC\lowercase{s} by duplicating ISS\lowercase{s}}
\label{induce}
%%%%%%%%%%%%%%%%%%%%%%%%%%%%%%%%%%%%%%%%%%%%%%%%%%%%%%%%%%%%%%%

\begin{figure}[!t]
  \centering
  \includegraphics*[width=0.98\columnwidth]{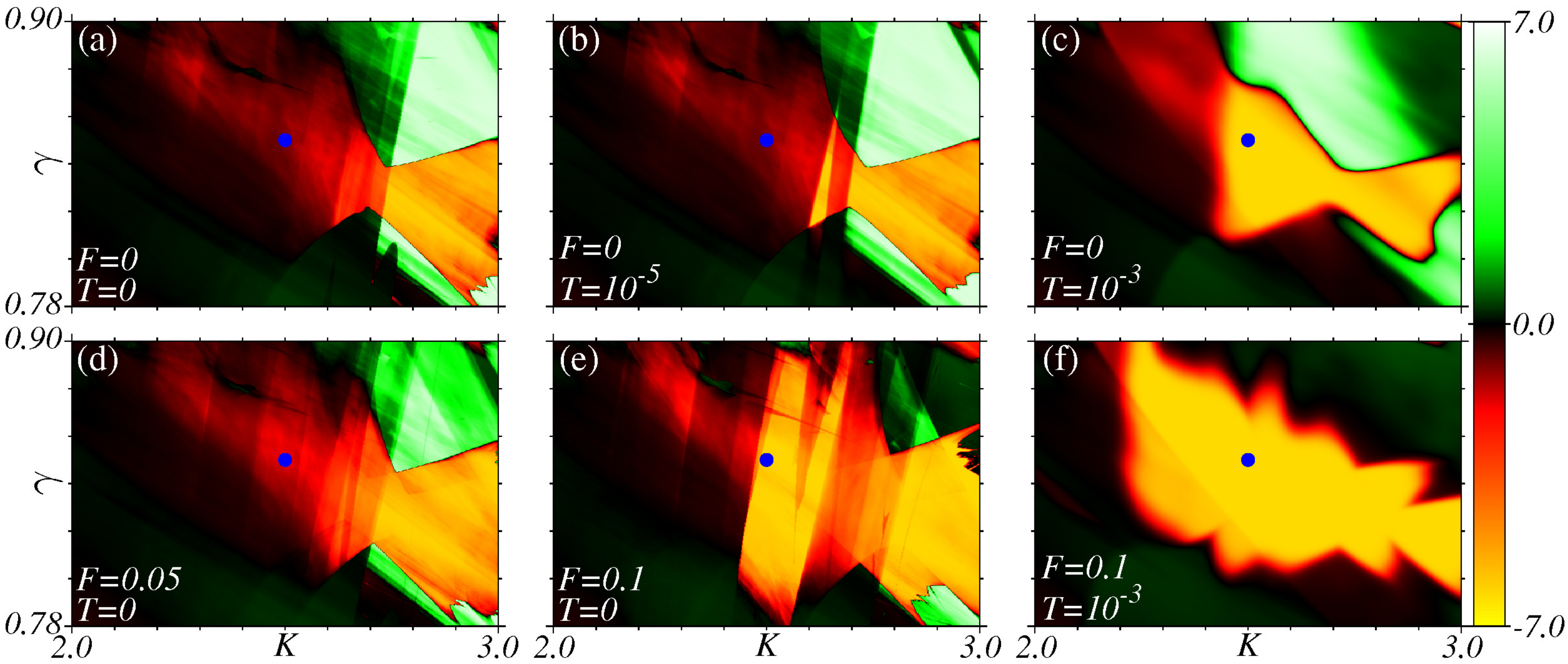}
  \caption{(Color online) RCs $\mathcal{J}$ plotted in the parameter
    space $(K,\gamma$) inside the interval
    $(K_{\mbox{\tiny min}}, K_{\mbox{\tiny max}})=(2.0,3.0)$ and
    $(\gamma_{\mbox{\tiny min}}, \gamma_{\mbox{\tiny max}})=(0.78,0.90)$.
    The values used for $F$ and $T$ are indicated in each panel and
    the blue point is located at $K=2.5$, $\gamma=0.85$, for all
    cases.}   
  \label{act}
\end{figure}

This section is devoted to discuss the effect of periodic
perturbations on the amazing phenomenon of {\it thermal activation},
triggered by the thermal bath coupled to the ratchet system. For the
RM (\ref{map-ratchet}) with $F=0$, there is a specific region in the
parameter space $(K,\gamma)$ where RCs are thermally activated instead
of being destroyed, as reported in \cite{mcb-1101141022013}. The RC
$\mathcal{J}$ in the vicinity of this region is plotted in
Fig.~\ref{act}(a) with $T=0$ and it is possible to identify part of a
non-cuspidal ISS \cite{ENDLER2006681} (see Figure $11$ in
\cite{Celestino2014139} for a detailed discussion). This kind of ISS,
together with cuspidal and shrimp-like ISSs, are generic patterns
which should appear in the parameter space of dissipative ratchet
system \cite{cmab-1062341012011}. Increasing the temperature to
$T=10^{-5}$ and $T=10^{-3}$, the effect of inducing RCs is observed in
Figs.~\ref{act}(b) and \ref{act}(c), respectively. Despite the {\it
  typical} effect of destruction of the ISSs by thermal effects, a
negative RC $\mathcal{J}$ is generated in regions around $K=2.5$,
$\gamma=0.85$, parameter combination represented by the blue
point. The thermal induced RC is due to a symmetry breaking of
periodic attractors in the momentum direction, as explained in
Ref.~\cite{mcb-1101141022013}.    
%
%%%%%%%%%%%%%%%%%%%%%%%%%%%%%%%%%%%%%%%%%%%%%%%%%%%%%%%%%%%%%%%%%%%%%%%%%
%%%%%%%%%%%%%%%%%%%%%%%%%%%%%%%%%%%%%%%%%%%%%%%%%%%%%%%%%%%%%%%%%%%%%%%%%

Besides the thermal activation of RCs, it is also possible to obtain
induced RCs when $F\neq 0$, even for $T=0$. However, the origin of
these induced RCs is slightly different, as explained below. In
Fig.~\ref{act}(d) and (e) a per-$2$ sequence $+F,-F,+F,-F,\ldots$ was
used with $F=0.05$ and $F=0.1$, respectively, and $T=0$ for both
cases. Comparing Fig.~\ref{act}(a) with Figs.~\ref{act}(d) and
\ref{act}(e), we conclude that, increasing $F$, a large region with
negative values of $\mathcal{J}$ (orange and yellow regions) around
the blue point emerges, {\it i.e.}, occurs the activation of optimal
ratchet currents. However, unlike the case where thermal effects are
present, the destruction of the non-cuspidal ISS is avoided when only
the external force $F$ is considered. In fact, this ISS is duplicated
(not clearly recognizable in Figs.~\ref{act}(d) and
\ref{act}(e)). Besides this, even if $T\neq 0$, using a per-$2$
external perturbation it is possible to induce negative RCs in a
larger portion of the parameter space, as displayed in
Fig.~\ref{act}(f). When comparing the percentage of occupied area
where $\mathcal{J}<-5$ in Fig.~\ref{act}(a) and Fig.~\ref{act}(c), a
gain of $109\%$ is obtained by using only thermal noise and,
comparing Figs.~\ref{act}(a) and \ref{act}(f), the  increasing area
is astonishing, around $359\%$ by adding the external  
perturbation $F_j$.

\begin{figure}[!h]
  \centering
  \includegraphics*[width=0.93\columnwidth]{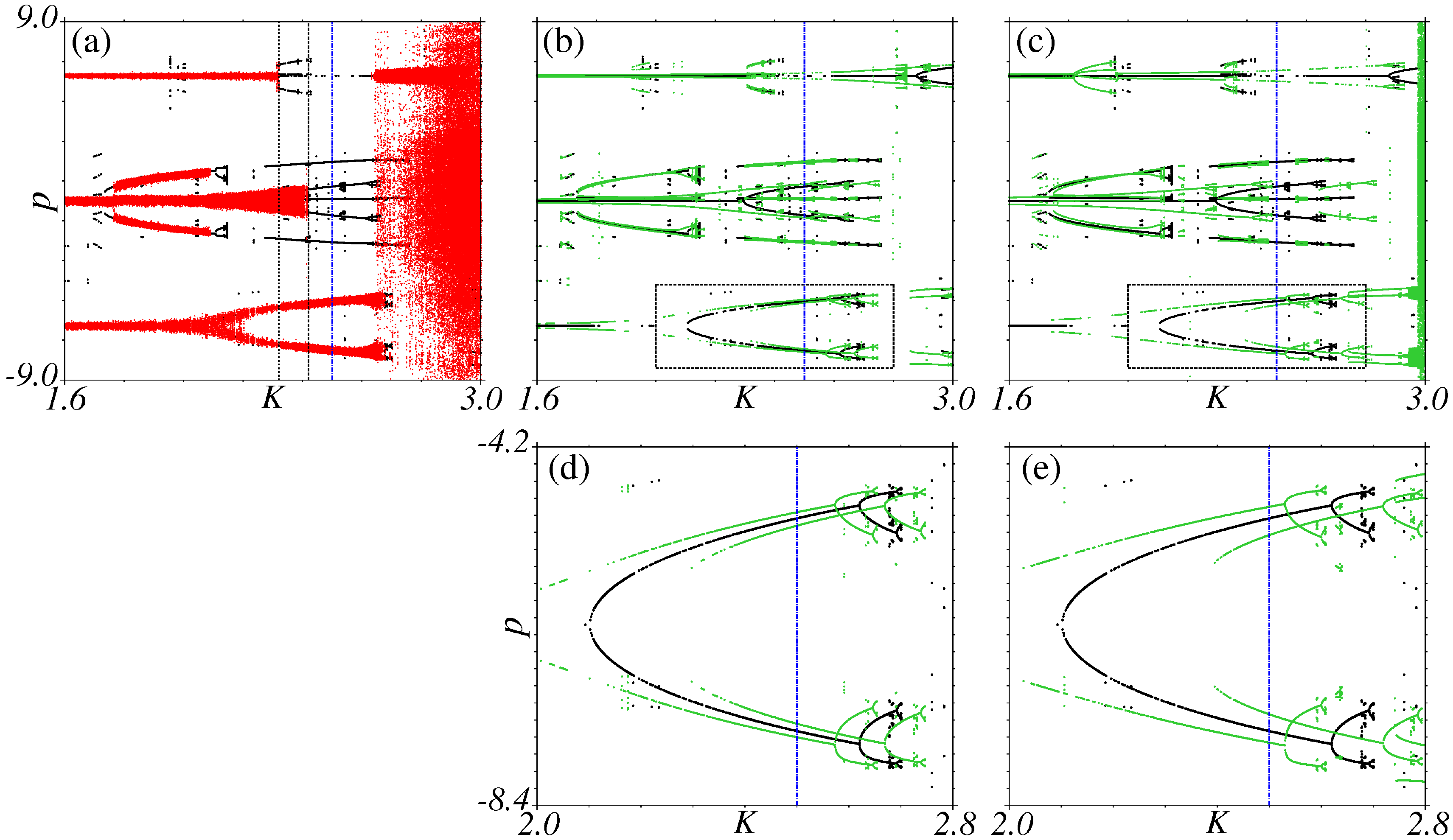}
  \caption{(Color online) Bifurcation diagram $(K,p)$ for
    $\gamma=0.85$ and
    $(K_{\mbox{\tiny min}}, K_{\mbox{\tiny max}})=(1.6,3.0)$ and
    $(p_{\mbox{\tiny min}}, p_{\mbox{\tiny max}})=(-9,9)$. In all
    panels, black dots represent the case $F=T=0$ and the vertical
    blue line corresponds to $K=2.5$. In (a), the case $T=10^{-3}$ and
    $F=0$ is plotted with red points and the vertical dotted
      and dashed black lines indicate the parameters $K=2.32$ and
      $2.42$, respectively. Keeping $T=0$, the cases with $F=0.05$
    and $F=0.1$ are displayed in (b) and (c), respectively, using
    green points. The panels (d) and (e) show the magnification of the
    black box in (b) and (c).} 
  \label{bif}
\end{figure}

To understand what happens when thermal effects and/or external
perturbations are considered in the RM (\ref{map-ratchet}), we plotted
in Fig.~\ref{bif} the bifurcation diagram $(K,p)$ for the dissipation
$\gamma=0.85$. In all panels the, black dots represent the cases for
which $F=T=0$. As reported in \cite{mcb-1101141022013} and
corroborated here in Fig.~\ref{bif}(a), for $T=10^{-3}$ and $F=0$ 
(red dots), the attractor located at $p=+2\pi$ is destroyed for
$K\approx 2.32$, value indicated by the vertical dotted black line in
Fig.~\ref{bif}(a). Moreover, for $K\approx 2.42$ [vertical   
dashed black line in Fig.~\ref{bif}(a)], all attractors symmetrically 
positioned around $p=0$ disappear, remaining only the per-$2$ 
attractor located  around $p=-2\pi$, leading to a negative RC
$\mathcal{J}=-2\pi$ in this region. Therefore, the vertical blue
line represents the parameter $K=2.5$, showing that for the parameter
combination $K=2.5$, $\gamma=0.85$ (see blue point in Fig.~\ref{act})
remains only an attractor at $p=-2\pi$ for $T=10^{-3}$, justifying the
thermal activation of RC.   

On the other hand, by considering $F\neq 0$ and $T=0$, case
represented in Fig.~\ref{bif} by the green dots, no attractor is
destroyed and the negative RC is induced by the birth of a {\it new}
attractor located around $p=-2\pi$ for $K\approx 2.29$, as we can see
in Fig.~\ref{bif}(b) and in its magnification in Fig.~\ref{bif}(d),
where an intensity $F=0.05$ was used. With the born of a new attractor
in the negative side of $p$, the RC $\mathcal{J}$ becomes negative
even if the attractor located at $p=+2\pi$ is still there. Increasing
the strength of $F$, the new independent bifurcation diagram begins to
move away as displayed in Figs.~\ref{bif}(c) and \ref{bif}(e) for
$F=0.1$, and the negative RC takes place in a larger range of $K$, as
is possible to note in Fig.~\ref{act}(e) for the same value of $F$. In
Figs.~\ref{bif}(d) and \ref{bif}(e), we can see that the birth of the
new attractor is due a saddle-node bifurcation, which follows exactly
the same recipe discussed in the Ref.~\cite{rafael17-1} for the
quadratic map and in \cite{rafael17-2} for the H\'enon map.  
%
%%%%%%%%%%%%%%%%%%%%%%%%%%%%%%%%%%%%%%%%%%%%%%%%%%%%%%%%%%%%%%%%%%%%%%%%%
%%%%%%%%%%%%%%%%%%%%%%%%%%%%%%%%%%%%%%%%%%%%%%%%%%%%%%%%%%%%%%%%%%%%%%%%%

%%%%%%%%%%%%%%%%%%%%%%%%%%%%%%%%%%%%%%%%%%%%%%%%%%%%%%%%%%%%%%%%%%%%%%%%%
\section{Summary and conclusions}
\label{conc}
%%%%%%%%%%%%%%%%%%%%%%%%%%%%%%%%%%%%%%%%%%%%%%%%%%%%%%%%%%%%%%%%%%%%%%%%%

In this work, a generic Ratchet Mapping (RM) which presents all
essential features regarding unbiased current [defined by
  Eq. (\ref{map-ratchet})] was submitted to a $k$-periodic external
perturbation $F_j$ to optimize domains with non-zero RC $\mathcal{J}$
in parameter spaces. In Sec.~\ref{meth}, we discuss a specif protocol
very useful to multiply and proliferate attractors and ISSs in phase
and parameter spaces, respectively. This allows us to enlarge stable
domains and, consequently, parameter combinations which contain
optimal values of RCs, even when inevitable thermal effects are
present. Following rules that relate the period $k$ of the external
perturbation and the period $m$ of attractors obtained for parameters
chosen inside {\it generic} ISSs, we are able to obtain an enlargement
in the area with negative RCs $\mathcal{J}$ of around $77.5\%$ for
$T=10^{-5}$, and $78.1\%$ for $T=10^{-3}$.

In other regions of the parameter space where no RCs are present,
using the periodic perturbation it is possible to induce optimal
RCs. In this case, the negative RC is induced by a saddle-node
bifurcation that generates a new attractor around $p=-2\pi$. The
advantage of generating RCs using periodic external perturbations is
that the per-$m$ ISSs with optimal RCs are enlarged when $\omega=m/k
\in \mathbb{Z}$, instead of being destroyed when thermal activation
are used \cite{mcb-1101141022013}. Even if thermal effects are
considered, the generation of RCs for $F\neq 0$ is more
efficient. Since any process in nature may be affected by intrinsic
thermal effects, these findings are of great interest from the
experimental point of view. Indeed, this finding is very interesting,
specially for transport problems related to experimental apparatus
with arbitrary low accuracy in nonlinearity parameters. Actually, we
hope that by applying the technique used in this work one is able to
experimentally reach unknown domains in parameter space of transport  
problems.  

%%%%%%%%%%%%%%%%%%%%%%%%%%%%%%%%%%%%%%%%%%%%%%%%%%%%%%%%%%%%%%%
\section*{Acknowledgments}
%%%%%%%%%%%%%%%%%%%%%%%%%%%%%%%%%%%%%%%%%%%%%%%%%%%%%%%%%%%%%%%
R.M.S. thanks CAPES (Brazil) and C.M. and M.W.B. thank CNPq (Brazil)
for financial support. C.M. also thanks FAPESC (Brazil) for financial
support. The  authors  also  acknowledge computational support from
Professor Carlos M.~de Carvalho at LFTC-DFis-UFPR.

%%%%%%%%%%%%%%%%%%%%%%%%%%%%%%%%%%%%%%%%%%%%%%%%%%%%%%%%%%%%%%%
%\section*{References}
%\bibliography{references}
%%%%%%%%%%%%%%%%%%%%%%%%%%%%%%%%%%%%%%%%%%%%%%%%%%%%%%%%%%%%%%%

\end{document}